\documentclass{PoS}

\usepackage{graphicx}% 
\usepackage{epsf}
\usepackage{epsfig}

\usepackage{amsmath}
\usepackage{amsfonts,amssymb}
\usepackage{bm}% bold math

\usepackage{dsfont}
\usepackage{slashed}
\usepackage{booktabs}
\usepackage{placeins}

\newcommand{\cl}{\centerline}

\newcommand{\be}{\begin{equation}}
\newcommand{\ee}{\end{equation}}
\newcommand{\bea}{\begin{eqnarray}}
\newcommand{\eea}{\end{eqnarray}}

\def\rmsmall #1{\mbox{\scriptsize #1}}

\newcommand{\Dlr}{\buildrel \leftrightarrow \over D\raise-1pt\hbox{}}
\newcommand{\msbar}{\overline{\rm MS}}

\title{Recent progress in hadron structure from Lattice QCD}

\ShortTitle{Recent progress in hadron structure from Lattice QCD}

\author{\speaker{Martha Constantinou} \\
        University of Cyprus \\ $\&$ \\ Computation-based Science and
        Technology Research Center, The Cyprus Institute \\
        E-mail: \email{marthac@ucy.ac.cy}}

\abstract{
We review recent progress in hadron structure 
using lattice QCD simulations, with main focus in the
evaluation  of nucleon quantities such as the axial and tensor
charges, and the spin content of the nucleon, using simulations
at pion masses close to the physical value. We  highlight
developments on the evaluation of the gluon moment, a new direct approach to compute quark
parton distributions functions on the lattice, as well as, the neutron electric dipole moment. A
discussion of the systematic uncertainties and the computation of the
disconnected contributions using dynamical simulations is also
included. 
}

\FullConference{The 8th International Workshop on Chiral Dynamics \\
                 29 June 2015 - 03 July 2015\\
                 Pisa, Italy}

\begin{document}

\section{Motivation}

Lattice QCD (LQCD) is a non-perturbative
approach that provides a powerful tool for  {\it ab initio} evaluation
of hadron observables. These include both quantities that are well
determined experimentally, but also those that are not  easily
accessible in experiment. Thus, LQCD provides input to
phenomenology and to searches  for beyond the Standard Model
Physics. Recent progress in the simulation of LQCD has been
impressive, mainly due to the improvements in the algorithms,
development of new techniques, and increase in computational
power. This enabled simulations to be carried out at parameters very
close to their physical values.

Understanding nucleon structure from first principles is considered a
milestone of hadron physics and numerous experiments have been
devoted to its study, starting with the measurements of the
electromagnetic form factors initiated more than 50 years
ago. Reproducing these key observables within the lattice QCD 
formulation is a prerequisite to obtaining reliable predictions on
observables that explore Physics beyond the Standard Model. There is a
rich experimental program in major facilities (CERN, JLab, MAMI, MESA, PSI,
JPARC, etc) investigating hadron structure, such as the proton radius,
electric dipole moments and scalar and tensor interactions.

The 12~GeV upgrade of the Continuous Electron Beam
Accelerator Facility  at JLab  will allow to employ new
methods for studying the basic properties of hadrons. Hadron structure has
been an essential part of the physics program, which involves new and
interesting high precision experiments, such as nucleon resonance
studies with CLAS12, the longitudinal spin structure of the nucleon,
meson spectroscopy with low momentum transfer electron scattering,
high precision measurement of the proton charge radius, and many more.

The experiments on the proton radius have attracted a lot of interest
since accurate measurements of the root mean square charge radius from muonic
hydrogen~\cite{Antognini} ($\langle r^2_p \rangle_{\mu H} = 0.84\,{\rm fm}^2$) 
is 7.7$\sigma$ yielded a value smaller that the radius determined from
elastic e-p scattering and hydrogen spectroscopy ($\langle r^2_p
\rangle_{e p} =0.88\,{\rm fm}^2$) \cite{Mohr:2012tt} (see
Ref.~\cite{Carlson:2015jba} for a review). The 4$\%$
difference in the two measurements is currently not explained. We note
that the measurements in the muonic hydrogen experiments are ten times
more accurate than other measurements and they are very sensitive to
the proton size. In particular, the radius is measured from the energy
difference between the 2P and 2S states of the muonic
hydrogen~\cite{Pohl2010} and more accurate experiments are planned at
PSI. 

The above few examples illustrate that hadron structure is a very rich
field of research relevant to new physics searches. Thus, lattice QCD
does not only provide input to on-going experiments, but also gives
guidance to new experiments within a robust theoretical framework. 

Being one of the building-blocks in the universe, the nucleon provides
an extremely valuable laboratory for studying strong dynamics
providing important input that can also shed light in new physics searches. 
Although it is the only stable hadron in the Standard Model,
its structure is not fully understood yet. There have been several
recent lattice QCD results on nucleon observables. In these
proceedings we discuss representative observables probing hadron
structure, as well as, challenges involved in their computations. Topics to
be covered include the nucleon axial and tensor charges, the nucleon
spin, including disconnected contributions, neutron electric dipole
moment, the first gluon moment of parton distribution functions  (PDFs), and a direct method for
computing quasi-distribution functions on the lattice. The systematic
uncertainties related to nucleon matrix elements are also investigated.

\section{Nucleon Matrix Elements}

\begin{figure}[!h]
\cl{\includegraphics[scale=0.5]{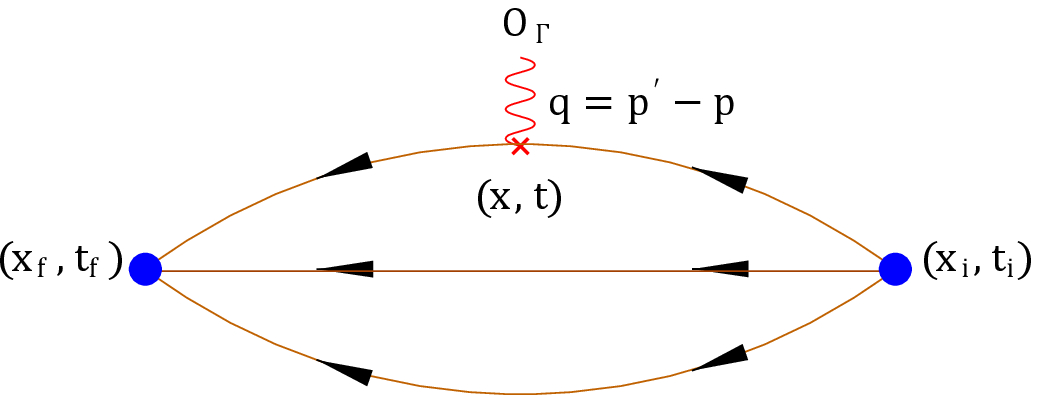} \hspace{1cm}
\includegraphics[scale=0.5]{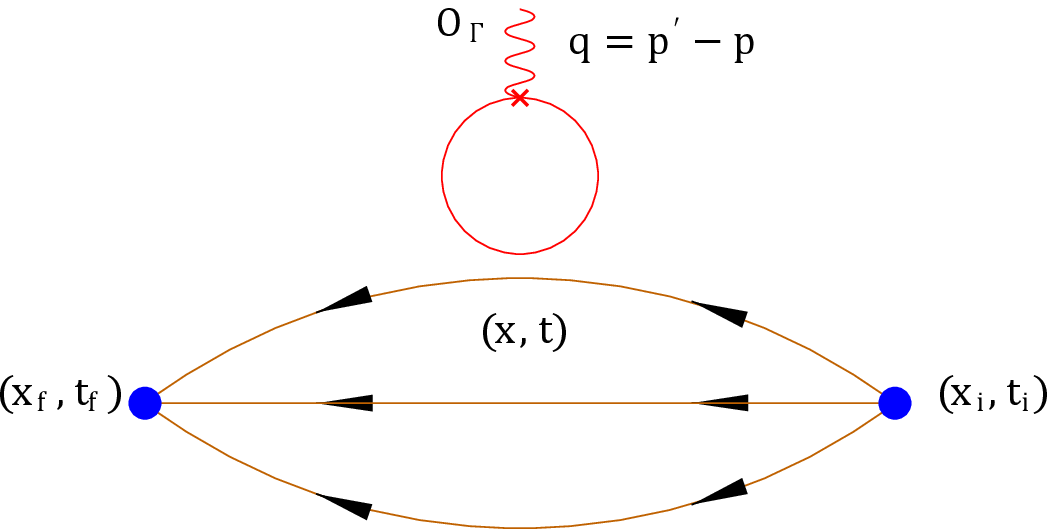}}
\vskip -0.1cm
\caption{Connected (left) and disconnected (right) contributions to
the nucleon three-point function.}
\label{fig1}
\end{figure}

In the evaluation of nucleon matrix elements in LQCD there are two type of diagrams
entering, shown in Fig.~\ref{fig1}. The disconnected diagram has been neglected
in most of the studies because it is very noisy and expensive to
compute. However, in the last few years a number of groups are
studying various techniques for its computation using dynamical simulations.
\vskip -0.2cm
\FloatBarrier
For the computation of nucleon matrix elements one constructs  
two- and  three-point correlation functions defined as 
\vspace{-0.15cm}
\bea
 G^{2pt}(\vec q, t_f) &=& \sum_{\vec x_f} \, e^{-i\vec x_f \cdot \vec q}\,
\Gamma^0_{\beta\alpha}\, \langle {{J_{\alpha}(\vec x_f,t_f)}}{{\overline{J}_{\beta}(0)}} \rangle \,, \\
 G^{3pt}_{\cal O}(\Gamma^\mu,\vec q, t_f) &=& \sum_{\vec x_f, \vec x} \, e^{i\vec x \cdot \vec q}\,e^{-i\vec x_f \cdot \vec p'}
     \Gamma^\mu_{\beta\alpha}\, \langle { J_{\alpha}(\vec x_f,t_f)} {\cal O}(\vec x,t) {\overline{J}_{\beta}(0)} \rangle\,,
\eea
\vskip -0.2cm
\noindent
appropriately projected in order to compute the quantities of
interest. For instance, the  projectors $\Gamma^\mu$ are usually defined as
$\Gamma^0 \equiv \frac{1}{4}(1+\gamma_0),\,\, \Gamma^k \equiv \Gamma^0\cdot\gamma_5\cdot\gamma_k\,$.
The lattice data are extracted from dimensionless ratio of the two-
and three-point correlation functions
%\vspace{-0.05cm}
\bea
R_{\cal O}(\Gamma,\vec q, t, t_f) \hspace*{-.2cm}&{=}&\hspace*{-.2cm} \frac{G^{3pt}_{\cal O}(\Gamma,\vec q,t)}{G^{2pt}(\vec 0, t_f)}
\hspace*{-0.1cm}\times\hspace*{-0.1cm}\sqrt{\frac{G^{2pt}({-}\vec q, t_f{-}t)G^{2pt}(\vec 0, t)G^{2pt}(\vec 0, t_f)}{G^{2pt}(\vec 0  , t_f{-}t)G^{2pt}({-}\vec q,t)G^{2pt}({-}\vec q,t_f)}}\,\,
{\rightarrow \atop {{t_f{-}t\rightarrow \infty} \atop {t{-}t_i\rightarrow \infty}}}\,\,
\Pi (\Gamma,\vec q) \,.
\label{EqRatio}
\eea
%\vskip -0.05cm
\noindent
The above ratio is considered optimized since it does not contain
potentially noisy two-point functions at large separations and also
correlations between its different factors reduce the statistical
noise. The most common method to extract the desired matrix element is to 
look for a plateau with respect to the current insertion time, $t$ (or,
alternatively, the sink time, $t_f$), which should be located at a
time well separated from the creation and annihilation times in order
to ensure single state dominance. To establish proper connection to
experiments we apply renormalization which, for most of the quantities
discussed in this review, is multiplicative
\vspace{-0.175cm}
\be
\Pi^R (\Gamma,\vec q) = Z_{\cal O}\,\Pi (\Gamma,\vec q)\,.
\ee
\vskip -0.175cm
The renormalized matrix elements can be parameterized in terms of
Generalized Form Factors (GFFs), and the decomposition follows the
symmetry properties of QCD. As an example we take the axial current 
insertion, which decomposes into two Lorentz invariant Form Factors
(FFs), the axial ($G_A$) and induced pseudoscalar ($G_p$)
\vspace{-0.3cm}
\be
\hspace{-0.4cm}
\langle N(p',s')|\bar\psi(x)\,\gamma_\mu\,\gamma_5\,\psi(x)|N(p,s)\rangle= i \Bigg(\frac{
            m_N^2}{E_N({\bf p}')E_N({\bf p})}\Bigg)^{1/2} \hspace{-0.25cm}
            \bar{u}_N(p',s') \Bigg[
            G_A(q^2)\gamma_\mu\gamma_5
            +\frac{q_\mu \gamma_5}{2m_N}G_p(q^2)
            \Bigg]u_N(p,s)\,,
\label{axial_decomp}
\ee
\vskip -0.1cm
\noindent
where $q^2$ is the momentum transfer in Minkowski space (hereafter, $Q^2=-q^2$).

Here, I will mostly consider the flavor isovector
combination for which the disconnected contribution cancels out;
strictly speaking, this happens for actions with  exact isospin
symmetry. Another advantage of the isovector combination is that 
the renormalization simplifies considerably.

\subsection{Systematic uncertainties}

The systematic uncertainties are important aspects of lattice
computations that need to be addressed carefully. In a nutshell, such
systematics are: 

\noindent
$\bullet$ cut-off effects due to the introduction of a finite lattice
spacing. For a proper continuum extrapolation one requires
simulations for, at least, three values of the lattice spacing, which is
computationally very costly, especially as we approach the physical
point. To minimize this systematic, gauge configurations are generated
 employing improved actions with a value of the lattice that ensures small or negligible  cut-off effects compared to the statistical accuracy.

\noindent
$\bullet$ finite volume effects due to the finite extent of the space-time box. These in general depend on the quantity under 
study. Ideally, simulations should be performed at multiple volumes, so
that the infinite volume limit can be taken. This requires significant
computer resources. As a rule of thumb one needs $L\,m_\pi$ larger than $3.5$
to suppress finite volume effects.

\noindent
$\bullet$ contamination from other hadron states due to the fact that
the interpolating field used to create a hadron of given quantum
numbers couple in addition to states higher in energy. While
for two-point functions identification of the lowest energy state is straight forward, 
for three-point functions it is more saddle, and there are various
methods to extract information from lattice data. The most common
approach is the so called plateau method in which one probes the large
Euclidean time evolution of the ratio in Eq.~(\ref{EqRatio})
\vspace{-.18cm}
\be
R_{\cal O}(\Gamma,t_i,t,t_f) {\rightarrow \atop {{(t_f{-}t)\,\Delta >>1} \atop {(t{-}t_i)\,\Delta >>1 }}}\,\,
{\cal M}\Bigg[ 1 {+} \alpha\,e^{-(t_f{-}t)\,\Delta(p')} {+} 
\beta\,e^{-(t{-}t_i)\,\Delta(p)} {+}\cdots \Bigg]\,.
\ee
\vspace{-.05cm}
In the above equation the excited states contributions
fall exponentially with the sink-insertion ($t_f-t$) and
insertion-source ($t-t_i$) time separation. So, it is possible to
reduce the unwanted excited states contamination by increasing the
source-sink separation, but this comes with a cost of increased
statistical noise.

Another method is the so-called summation method in which we sum the
ratio from the source to the sink, and thus, the excited state
contaminations are suppressed by exponentials decaying with 
$(t_f-t_i)$ rather than $(t_f - t)$ and $(t - t_i)$.  
However, one needs the slope of the summed ratio
\vspace{-.065cm}
\be
\sum_{t=t_i}^{t_f}\,R(t_i,t,t_f) = {\rm const.} +
{\cal M}\, (t_f-t_i) + 
{\cal O}\Bigg( e^{-((t_f-t_i)\,\Delta(p'))}\Bigg) + 
{\cal O}\Bigg( e^{-((t_f-t_i)\,\Delta(p))} \Bigg) \,.
\ee
\vskip -.065cm

\noindent
$\bullet$ simulations at unphysically large values of the pion mass
due to limitations on the computational resources and optimization
techniques. Then one typically uses chiral perturbation theory
($\chi$PT) to carry the extrapolation to the physical point, with low
energy constants determined by over-constraining the fits using
experimental, as well as, lattice data. Over the last years
simulations at physical parameters have become feasible, which can be
compared directly to experimental and phenomenological data. This is a
substantial step forward since the chiral extrapolation to the
physical point is avoided, which is often difficult and can lead to
rather large systematic uncertainties, in particular in the baryon sector.

\noindent
$\bullet$ renormalization, which might involve mixing with other
observables. In addition the data should be converted to the $\msbar$
scheme in order to be compared to experimental and phenomenological
data. This conversion is performed using perturbative expressions to
finite order in the coupling constant and this might bring in
systematic uncertainties; using higher-loop expressions (typically
${\cal O}(g^6)$) exhibit very small systematics. More importantly,
renormalization functions computed non-perturbatively may carry
lattice artifacts, which can be removed by subtracting them utilizing
perturbation theory~\cite{Constantinou:2010gr,Constantinou:2014fka,Alexandrou:2015sea}. 

\section{Nucleon Charges}

\subsection{Axial Charge}

One of the fundamental nucleon observables is the axial charge, 
$g_A \equiv G_A(0)$, which is determined from the  forward matrix
element of the axial current, and gives the intrinsic quark spin
in the nucleon. It governs the rate of $\beta$-decay and has been
measured precisely. In LQCD the axial charge can be determined directly
from the evaluation of the matrix element and thus, there is no
ambiguity associated to fits. For this reasons, $g_A$ is an optimal
benchmark quantity for hadron structure computations, and it is
essential for LQCD to reproduce its experimental value or if a
deviation is observed to understand its origin.
\vspace*{-0.1cm}
\begin{figure}[!h]
\cl{\includegraphics[scale=0.38]{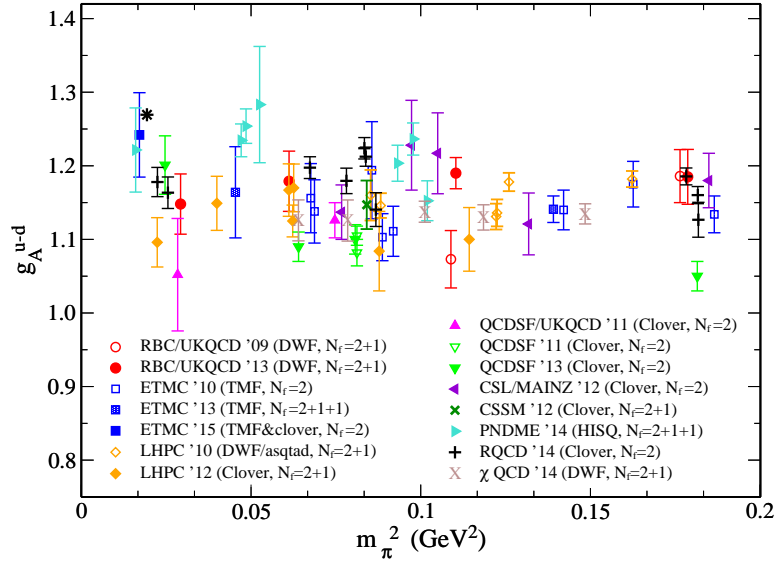}}
\vspace*{-0.35cm}
\caption{Collection of lattice results for $g_A$ corresponding to: $N_f{=}2{+}1$ DWF
(RBC/UKQCD~\cite{Yamazaki:2008py, Yamazaki:2009zq},
RBC/UKQCD~\cite{Ohta:2013qda}, $\chi$QCD~\cite{chiQCD14}),
$N_f{=}2{+}1$ DWF on asqtad sea (LHPC~\cite{Bratt:2010jn}), $N_f{=}2$
TMF (ETMC~\cite{Alexandrou:2010hf}), $N_f{=}2$ Clover
(QCDSF/UKQCD~\cite{Pleiter:2011gw}, CLS/MAINZ~\cite{Capitani:2012gj},
QCDSF~\cite{Horsley:2013ayv}, RQCD~\cite{Bali:2014nma}),
$N_f{=}2{+}1$ Clover (LHPC~\cite{Green:2012rr},
CSSM~\cite{Owen:2012ts}), $N_f{=}2{+}1{+}1$ TMF
(ETMC~\cite{Alexandrou:2013joa}), $N_f{=}2{+}1{+}1$ HISQ
(PNDME~\cite{Bhattacharya:2013ehc,PNDME14}), $N_f{=}2$ TMF with Clover
(ETMC~\cite{Abdel-Rehim:2015owa}). The asterisk is the experimental value~\cite{PDG12}.}
\label{fig_gA}
\end{figure}
\FloatBarrier
There are numerous computations of $g_A$ from many collaborations
and selected results are shown in Fig.~\ref{fig_gA} as a function
of $m_\pi^2$. These results have been obtained using dynamical 
gauge field configurations with ${\cal O}(a)$-improved lattice QCD
actions, namely Domain Wall Fermions (DWF), Hybrid, Clover,
Twisted Mass Fermions (TMF) and HISQ fermions (see caption of
Fig.~\ref{fig_gA} for references). For a meaningful comparison we
include only results obtained from the plateau method without any
volume corrections. The latest achievement of the Lattice Community
are the results at the physical point for which there is no necessity
of chiral extrapolation eliminating an up to now uncontrolled
extrapolation. The ones at the two lowest values of the pion
mass correspond to PNDME (128~MeV)~\cite{PNDME14} and ETMC
(133~MeV)~\cite{Abdel-Rehim:2015owa}, and are in agreement with the
experimental value: $g_A^{\rm exp}=0.2701(25)$~\cite{PDG12}. 
Of course the statistical errors are still large and it is necessary
to increase the statistics and study the volume and lattice spacing
dependence before finalizing these results. In addition, the results
shown in  Fig.~\ref{fig_gA} are at a given lattice spacing and volume
and, thus, systematic effects should be investigated. 

In summary, based on current results on the axial charge~\cite{Constantinou:2014tga},
we conclude that cut-off effects are small, at least for $a
\le 0.1$ fm, and no indication of significant excited state
contamination has been observed indicating that sink-source time
separation of about 1~fm is sufficient. No clear conclusion can be
extracted regarding finite volume effects that need further
investigation. It is worth stressing, however, that the value of $g_A$
determined close to the physical point by ETMC with $L\,m_\pi \sim 3$
($a<0.1$~fm), and by PNDME with $L\,m_\pi \sim 3.75$ ($a<0.1$~fm)
are in agreement with the experimental data.

We note that all high statistics studies of systematic uncertainties
have been performed at relatively large values of the pion mass. It is
thus essential to also perform similar investigations at values of the
pion mass closer to the physical one. Given that the signal to noise
error decreases exponentially as the pion mass decreases
one needs to increase considerably the number of independent
measurements leading to increase computational cost, and thus, noise
reduction methods are highly valuable.
%\be
%\sim \sqrt(N_{\rm meas})\times e^{-\left(m_N + 3\,m_\pi/2 \right)}
%\ee
%($N_{\rm meas}$: number of independent measurements, $m_N$: nucleon
%mass, $m_\pi$: pion mass)  

\subsection{Tensor Charge}

The nucleon scalar and tensor charges have not been studied in LQCD as extensively as 
$g_A$
since the contributions of effective
scalar and tensor interactions in the Standard Model are very small
(per-mil level). These interactions correspond to the non $V-A$
structure of weak interactions and serve as a test for new physics. 
Ongoing experiments using ultra-cold
neutrons~\cite{Abele:2002wc,Nico:2009zua,Young:2014mxa,Baessler:2014gia},
as well as planned ones~\cite{Bhattacharya:2011qm} will reach the
necessary precision to investigate such interactions.

In order to study the scalar and tensor interactions we add a term in
the effective Hamiltonian corresponding to new BSM physics, 
\vspace*{-0.2cm}
\be
H_{eff} = G_F \left( J_{V_A}^{l}\times J_{V_A}^{q} + \sum_i \epsilon_i {\cal O}_i^l \times {\cal O}_i^q \right)
\ee
\vskip -0.2cm
\noindent
where the sum includes operators with novel structure, such as the
scalar and tensor, which come with low-energy couplings that are
related to masses of new TeV-scale particles. 

Experimentally, bounds on the tensor coupling constant arise
in the radiative pion decay $\pi \to e \nu\gamma$, while new
experiments at Jefferson lab are planned using polarized 3He/Proton
aim at increasing its accuracy by an order of magnitude~\cite{Gao:2010av}. 
Also, experiments at LHC are expected to increase the limits to
contributions from tensor and scalar interactions by an order of
magnitude, making these observables interesting probes of new
physics. Computations of the scalar charge will also provide input for
dark matter searches, since experiments aiming at a direct detection
of dark matter, are based on measuring the recoil energy of a nucleon
hit by a dark matter candidate. In several supersymmetric scenarios
\cite{Ellis:2010kf} the dark matter nucleon interaction is mediated
through a Higgs boson. In this case the theoretical expression of the
spin independent scattering amplitude at zero momentum transfer
involves the quark content of the nucleon or the nucleon
$\sigma$-term, which is closely related to the scalar charge. Thus,
computations of the nucleon scalar and tensor charges within LQCD will
provide useful input for the ongoing experimental searches for BSM
physics.

\begin{figure}[!h]
\cl{\includegraphics[scale=0.38]{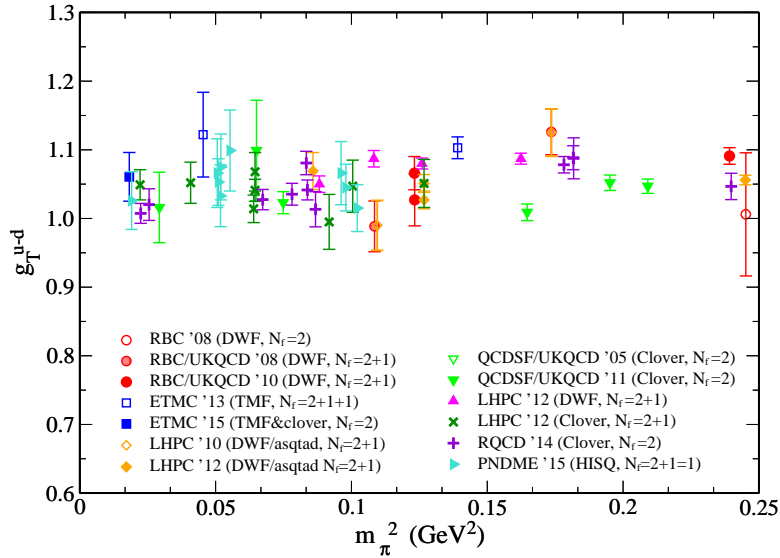}}
\caption{Lattice results for $g_T$ as a function of the $m_\pi^2$, corresponding to: 
$N_f{=}2$ Clover (QCDSF/UKQCD~\cite{Gockeler:2005cj,Pleiter:2011gw,Bali:2014nma}),
$N_f{=}2$ DWF (RBC~\cite{Lin:2008uz}),
$N_f{=}2{+}1$ DWF (RBC/UKQCD~\cite{Ohta:2008kd,Aoki:2010xg}, LHPC~\cite{Green:2012ej}),
$N_f{=}2{+}1$ DWF on asqtad sea (LHPC~\cite{Bratt:2010jn,Green:2012ej}), 
$N_f{=}2{+}1$ Clover (LHPC~\cite{Green:2012ej}),
$N_f{=}2{+}1{+}1$ TMF (ETMC~\cite{Alexandrou:2013wka}), 
$N_f{=}2{+}1{+}1$ HISQ (PNDME~\cite{Bhattacharya:2015wna}), 
$N_f{=}2$ TMF with Clover (ETMC~\cite{Abdel-Rehim:2015owa}).} 
\label{fig_gT}
\end{figure}

In these proceedings we focus on the tensor charge, which is the zeroth
moment of the transversity distribution functions, the last part among
the three quark distributions at leading twist
\vspace*{-0.2cm}
\be
\langle N(p',s')|\,{\cal O}_{T^\alpha}^{\mu\nu}\,|N(p,s)\rangle \,,\quad
{\cal O}_{T^\alpha}^{\mu\nu} = \bar{\psi}\,\sigma^{\mu\nu} \frac{\tau^\alpha}{2}\,\psi\,.
\ee
\vskip -0.2cm
\noindent
Results on the isovector tensor charge are compared in Fig~\ref{fig_gT} 
for several discretizations, lattice spacings, and volumes. For all
results, the plateau method has been chosen for a meaningful
comparison. Overall, there is a very good agreement among lattice data, which
also exhibit very mild pion mass dependence. We would like to highlight
the data at the physical point~\cite{Abdel-Rehim:2015owa,Bhattacharya:2015wna}
that provide a prediction for this quantity, free of uncontrolled
systematics due to chiral extrapolations.
\FloatBarrier
The excited states contamination for $g_T$ have been also investigated, 
revealing a weak dependence on the source-sink time separation, $T_{sink}$ 
\cite{Bali:2014nma,Bhattacharya:2015wna,Abdel-Rehim:2015owa}:
the values from the plateau method do not vary as a function of
$T_{sink}$ and are in agreement with the value extracted from the
summation method, within statistical uncertainties.

On the experimental side there are available data for $g_T$ obtained
from combined global analysis of the measured azimuthal asymmetries in
SIDIS and in $e^+ e^- \to h_1 h_2 X$ (see, e.g. \cite{Anselmino:2013vqa}).
There are also results from model predictions, for instance the
predictions by a covariant quark-diquark model. However, direct
comparison of the tensor charges from different models and scales is
not always meaningful, since the tensor charges are strongly scale-dependent
quantities, in particular at low values of the renormalization scale.
Therefore, ab initio calculations of $g_T$ from Lattice QCD are
extremely useful in providing reliable and model-independent predictions.

\section{PDFs on the Lattice}

Measurements of parton distribution functions  in high-energy
processes such as deep-inelastic lepton scattering and Drell-Yan in
hadron-hadron collisions are very interesting since they provide
information on the quark and gluon structure of a hadron. To leading
twist, these quantities give the probability of finding a specific
parton in the hadron carrying certain momentum and spin, in the
infinite momentum frame. Due to the fact that PDFs are light-cone 
correlation functions (quark and gluons fields are separated along the
light-cone, defined in the real Minkowski time), what is
calculated in LQCD are  Mellin moments expressed in terms of hadron matrix elements
of local operators. Although there is intense activity on the
computation of such moments in lattice QCD, it is highly desirable to
have information on the PDFs themselves. The Mellin moments are
related to the original PDFs through the operator product expansion
(OPE). However, the reconstruction of the PDFs appears to be
unfeasible since the signal-to-noise ratio becomes very small for
higher moments. Also for moments with more than 3 derivatives there is
unavoidable mixing with lower dimension operators, which complicates
the renormalization procedure. In addition, there is limited progress
in calculations of gluon moments (see Section~\ref{secGluonMoment}),
which require a disconnected insertion, has low signal quality and
operator mixing. 

Recently, a novel direct approach has been proposed by
Ji~\cite{Ji:2013dva}, suggesting that one can compute a parton
quasi-distribution function, $\tilde{q}(x, \Lambda, P, \Gamma)$, where
$x=k/|\vec{P}|$, $\Lambda$ is an ultraviolet cutoff scale, $\vec{P}$
is the momentum of the nucleon, and $\Gamma$ is the Dirac structure of
the operator under study. $\tilde{q}(x, \Lambda, P)$ is accessible on
the lattice, and for large momenta, one can establish connection with
the PDFs through a matching procedure. Such a matching appears in
one-loop perturbation theory~\cite{Xiong:2013bka} and the computation
of quasi-distribution functions has been carried out in
Ref.~\cite{Lin:2014zya} using $N_f{=}2{+}1{+}1$ HISQ gauge ensembles
with clover valence quarks, and more recently in
Ref~\cite{Alexandrou:2015rja} for $N_f{=}2{+}1{+}1$ twisted mass
fermions.

The momentum-dependent non-local static correlation is written as 
\be
\tilde{q}(x,\Lambda,P_z,\Gamma)  = \int_{-\infty}^{infty} \frac{dz}{4\pi} e^{-izk} \times
\left\langle P \right\vert \bar{\psi}(z)\Gamma e^{i g\int_0^z 
A_z(z^\prime) dz^\prime} \psi(0) \left\vert P\right\rangle \,,
\label{qtildeDef}
\ee
where $x$ is the momentum distribution and $e^{ig\int_0^z A_z(z^\prime)
  dz^\prime}$ is the Wilson line introduced to ensure gauge invariance
in the quark distribution. Also, for simplicity, the momentum
$\vec{P}$ is taken in the $z$-direction. One of the characteristics of
$\tilde{q}$ is that, unlike the case of the physical PDFs, is non=zero for $|x|>1$. When the momentum approached infinity one
recovers the physical distribution functions, $q(x,\mu)$, with the
infrared region remaining the same. For finite
but large enough momenta, $\tilde q$ and $q$ are related via
\bea
\label{q}
q(x,\mu) = q_{bare}(x)\left\{1 + \frac{\alpha_s}{2\pi} Z_F(\mu)
\right\} + \frac{\alpha_s}{2\pi}\int_x^1 q^{(1)} (x/y,\mu )
q_{bare}(y) \frac{dy}{y} + \mathcal{O}(\alpha_s^2)\,,\\
\tilde{q} (x,\Lambda,P_z) = q_{bare}(x)\left\{1 +
\frac{\alpha_s}{2\pi} \tilde{Z_F}(\Lambda,P_z) \right\} +
\frac{\alpha_s}{2\pi}\int_{x/x_c}^1 \tilde{q}^{(1)} (x/y,\Lambda,P_z )
q_{bare}(y) \frac{dy}{y} + \mathcal{O}(\alpha_s^2)\,,
\label{qtilde}
\eea
where $q_{bare}$ is the bare distribution, $Z_F$ and $\tilde Z_F$ are
the wave function corrections, $q^{(1)}$ and $\tilde{q}^{(1)}$ are the
vertex corrections. Also, $\mu$ is the renormalization scale and
$x_c=\Lambda/P_x$ is the largest possible value of the momentum
distribution, $x$. The leading UV divergences to the
quasi-distribution functions are computed in perturbation theory by
having $P_z$ fixed, while sending $\Lambda \to \infty$. The UV
regulator $\Lambda$ will be set to the renormalization scale $\mu$
when relating $\tilde{q}$ at finite momentum to $\tilde{q}$ at
infinite momentum. 

On the lattice one computes $\tilde{q}(x,\Lambda,P_z,\Gamma)$ as defined 
in Eq.~(\ref{qtildeDef}), which is then used in the lhs of Eq.~(\ref{qtilde}) 
to calculate the rhs of Eqs.~(\ref{q}) - (\ref{qtilde}) in order to
extract the quark distribution. This makes use of perturbation theory
and to date results exist for the non-singlet case and to ${\cal O}(a_s)$ 
for the vertex corrections and the self-energy. Thus, a combination of
Eqs.~(\ref{q}) - (\ref{qtilde}) leads to
\bea
\tilde{q} (x,\Lambda,P_z) &=& q(x, \mu) + \frac{\alpha_s}{2\pi}
q(x,\mu) \left\{\tilde{Z_F}(\Lambda,P_z)-Z_F(\mu) \right\}
\nonumber \\
&+& \frac{\alpha_s}{2\pi}\int_{x/x_c}^1 \left(\tilde{q}^{(1)}
(x/y,\Lambda,P_z )-q^{(1)} (x/y,\mu )\right) q(y, \mu) \frac{dy}{y} +
\mathcal{O}(\alpha_s^2) \,,
\label{qtilde1loop}
\eea
From the above expression one can isolate $q(x,\mu)$, and by including at
the same time anti-quarks ($\bar{q}(x)=-q(-x)$), Eq.~(\ref{qtilde1loop})
can be rewritten as
\bea
q(x, \mu) &=& \tilde{q} (x,\Lambda,P_z) - \frac{\alpha_s}{2\pi}
\tilde{q} (x,\Lambda,P_z) \delta  \left(\tilde{Z_F}(\Lambda,P_3)-Z_F(\mu)\right) \nonumber \\
& & - \frac{\alpha_s}{2\pi}\int_{-1}^1
\left(\tilde{q}^{(1)} (\xi,\Lambda,P_3 )- q^{(1)} (\xi,\mu ) \right) \tilde{q} (y,\Lambda,P_z) \frac{dy}{|y|} +
\mathcal{O}(\alpha_s^2)\,.
\eea
A nucleon mass correction in $M_N/P_z$ can be also made to an
arbitrary order. More details are given in Refs.~\cite{Lin:2014zya,Alexandrou:2015rja}.

A first study appeared in Ref~\cite{Lin:2014zya} for the unpolarized
and polarized quasi-distribution functions using clover valence
fermions on an $N_f{=}2{+}1{+}1$ ensemble of HISQ quarks corresponding
to $m_\pi\sim$310 MeV. The authors apply HYP smearing to the gauge
links, which appears to minimize the discretization effects. In
addition, since the multiplicative renormalization of $\tilde{q}$ has
not been computed yet, the smearing is important because it shift the
renormalization functions close to unity.

\begin{figure}[!h]
\cl{\includegraphics[scale=0.32]{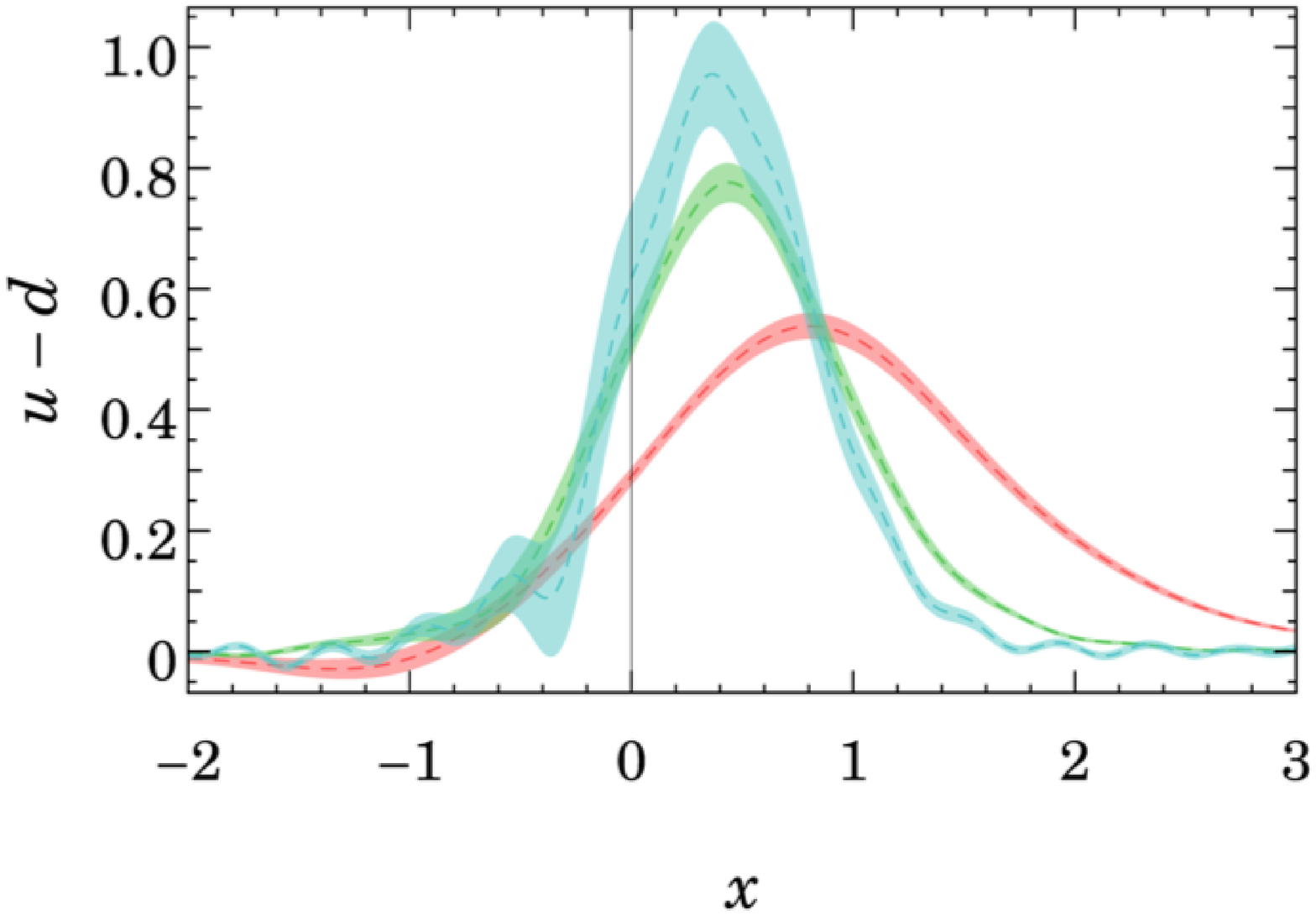}\,\,
\includegraphics[scale=0.33]{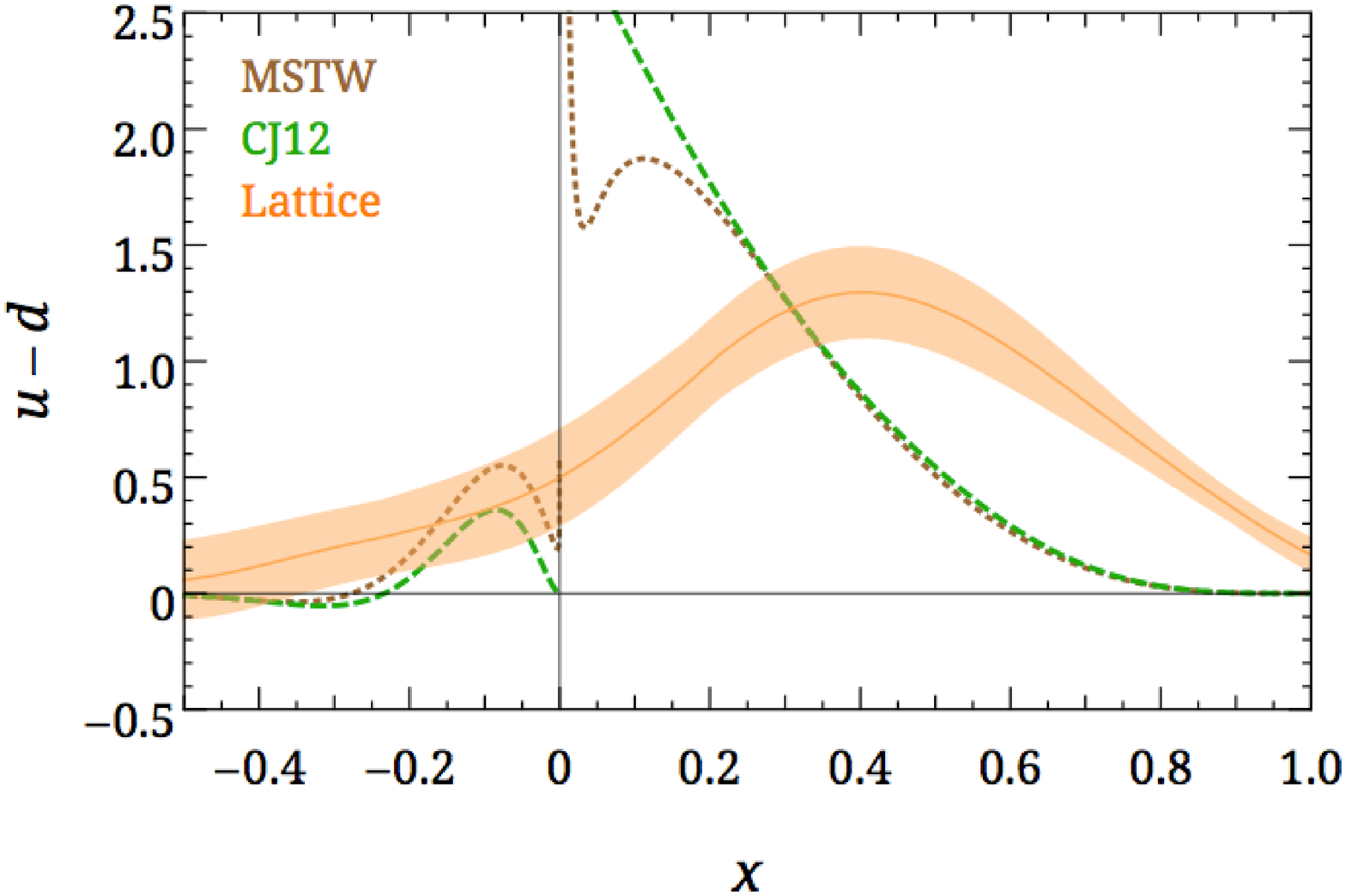}}
\vskip -0.4cm
\caption{Left: Lattice results for the isovector $\tilde{q}(x)$ using
momenta $P_z$: $\frac{2\pi}{L}$ (red), $\frac{4\pi}{L}$
(green),$\frac{6\pi}{L}$ (cyan). Right: Unpolarized isovector
$\tilde{q}(x)$ upon extrapolation in $P_z$ with 68$\%$ C.L (orange
band). Brown (green) dotted line corresponds to global analysis of
MSTW~\cite{Martin:2009iq} (CTEQ-JLab~\cite{Owens:2012bv}).}
\label{figPDFsMILC}
\end{figure}

For the unpolarized case, the matrix element calculated on the lattice is
\vspace*{-0.2cm}
\be
h(z,\Lambda,P_z) = \left\langle \vec{P}\right\vert \bar{\psi}(z) \gamma_3
\left( \prod_n U_z(n\hat{z})\right) \psi(0) \left\vert
\vec{P}\right\rangle \,.
\label{hPDFs}
\ee
\vskip -0.2cm
\noindent
Note that, in order to study the polarized operator one should simply
substitute the Dirac structure $\gamma_3$ with $\gamma_5\,\gamma_3$.
The matrix elements are computed with the nucleon boosted with
momentum $P_z=\frac{2\pi}{L},\,\frac{4\pi}{L},\,\frac{6\pi}{L}$. In
the left panel of Fig.~\ref{figPDFsMILC} we show results for the
isovector $\tilde{q}$ which is the Fourier transformation of
$h(z,\Lambda,P_z)$ in the $z$ coordinate  
\vspace*{-0.2cm}
\be
\tilde{q}_{\text{lat}}(x,\Lambda,P_z) = \int \frac{dz}{4\pi} e^{-izk}h(z,\Lambda,P_z)\,.
\ee 
\vskip -0.2cm
\noindent
The values for the momentum $P_z=\frac{2\pi}{L},\,\frac{4\pi}{L},\,\frac{6\pi}{L}$
are shown with a red, green and cyan color, respectively. One observation 
is that at momentum $\frac{2\pi}{L}$ the peak of the distribution is
centered around $x=1$, where the physical distribution is, in fact,
zero. For larger momenta the peak of the band moves towards zero and
centered around $x=0.5,\,0.4$ for $P_z=\frac{4\pi}{L},\,\frac{6\pi}{L}$, 
respectively. Also, the value of $\tilde{q}$ at $x=1$ reduces as the
momentum increases, which is expected, since for asymptotically large
momenta the quasi-distribution functions approach the physical ones.
In lattice calculations the maximum momenta are limited by statistical
accuracy, and thus, a large-momentum effective field theory should be
utilized to relate the finite-momentum $\tilde{q}$ to the physical ones.
To do so, one can use perturbative corrections, available to 1-loop
level~\cite{Xiong:2013bka}. Due to the fact that the momenta are not
infinitely large, the nucleon-mass corrections -expressed in terms of
$M_N^2/(4P_z^2)$- are also important and should taken into account as
explained in Ref.~\cite{Lin:2014zya}. The lattice data for $\tilde{q}$
upon the mass and 1-loop corrections have a milder momentum dependence, 
and one should finally extrapolate to the infinite momentum limit,
using the fit $a + b/P_z^2$. The extrapolated unpolarized isovector
$\tilde{q}$ is shown in the right panel of Fig.~\ref{figPDFsMILC},
where one observes that outside the region $|x| \ge 1$ the curve
drops significantly, as expected. The lattice data are plotted with
results from global analysis by MSTW~\cite{Martin:2009iq} and 
CTEQ-JLab~\cite{Owens:2012bv}, although no attempt for comparison is
made since the available lattice data cannot be extrapolated to the
chiral limit nor the continuum, and no source of systematics has been
addressed. 

\FloatBarrier

Another recent study of the unpolarized quasi-distribution function
was performed by ETMC, using an ensemble of $N_f{=}2{+}1{+}1$ twisted
mass fermions at $m_\pi\sim 373$MeV. The matrix elements are computed
for the 3 lowest momenta:
$P_z=\frac{2\pi}{L},\,\frac{4\pi}{L},\,\frac{6\pi}{L}$, since the
statistical noise does not allow to explore higher momenta. The
dependence on the source-sink separation is also studied, showing
compatibility within error bars. Two to five HYP smearing steps are
applied to the gauge links of the operator, which as mentioned above,
it is expected to bring the renormalization function closer to unity.
Since the renormalization function for this quantity is not yet
available (a perturbative calculation is in progress~\cite{ZquasiPDF}), 
a comparison between the unsmeared results with the data from various
HYP smearing steps, reveals the influence of the renormalization; it
appears that the smearing affect is stronger in the imaginary
part. However, the authors of Ref.~\cite{Alexandrou:2015rja} use the
renormalization function of the ultra-local vector current, since for
$z=0$, the operator reduces to the local vector current. As a test,
one can check the value of the renormalized $h(z=0,\Lambda,P_z)$,
defined in Eq.~\ref{hPDFs}, which is expected to be equal to 1. The
authors find for instance that $h^R(z=0,\Lambda,4\pi/L)=0.99(3)$.

\begin{figure}[!h]
\cl{\includegraphics[scale=0.31]{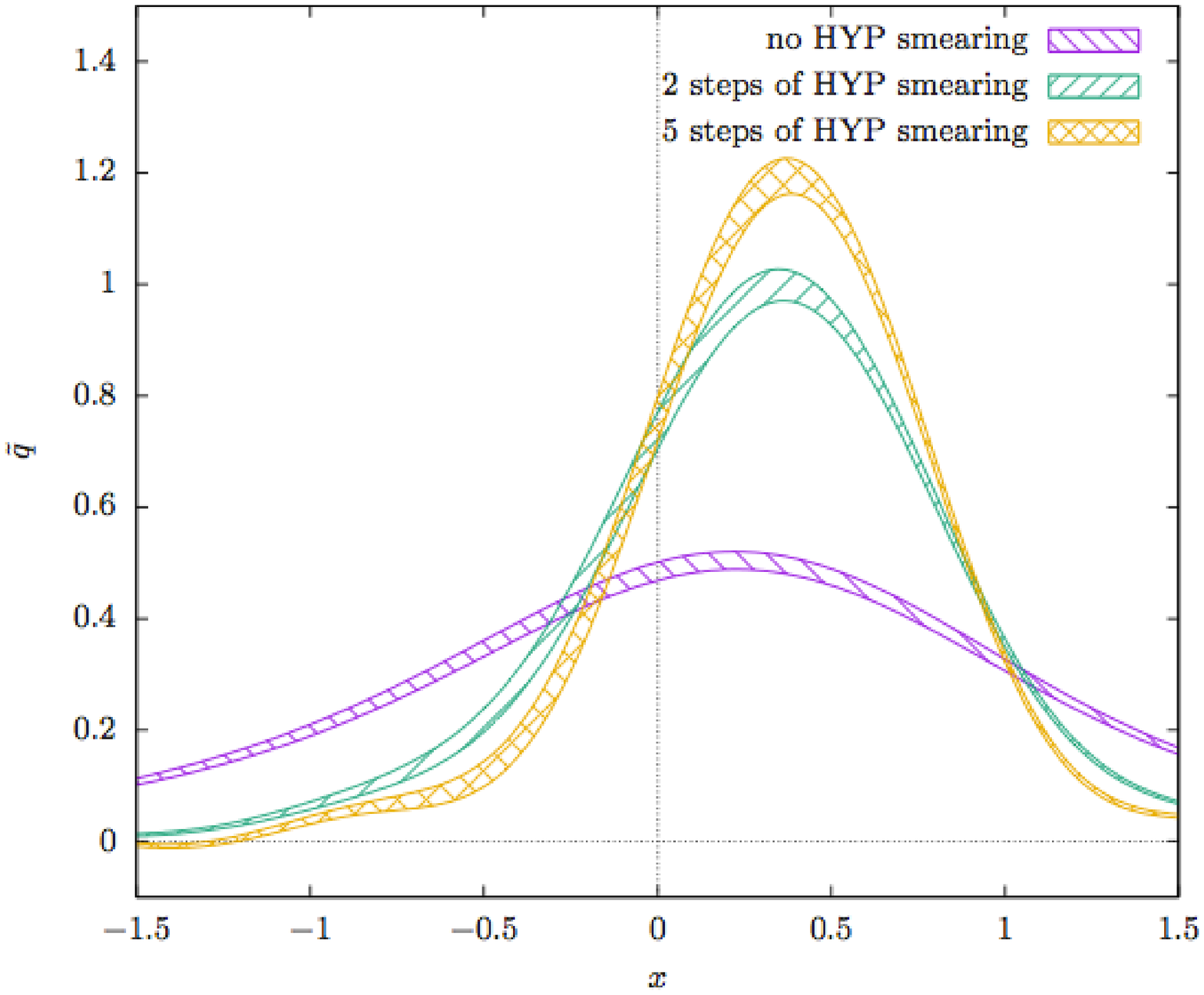}\,\,
\includegraphics[scale=0.29]{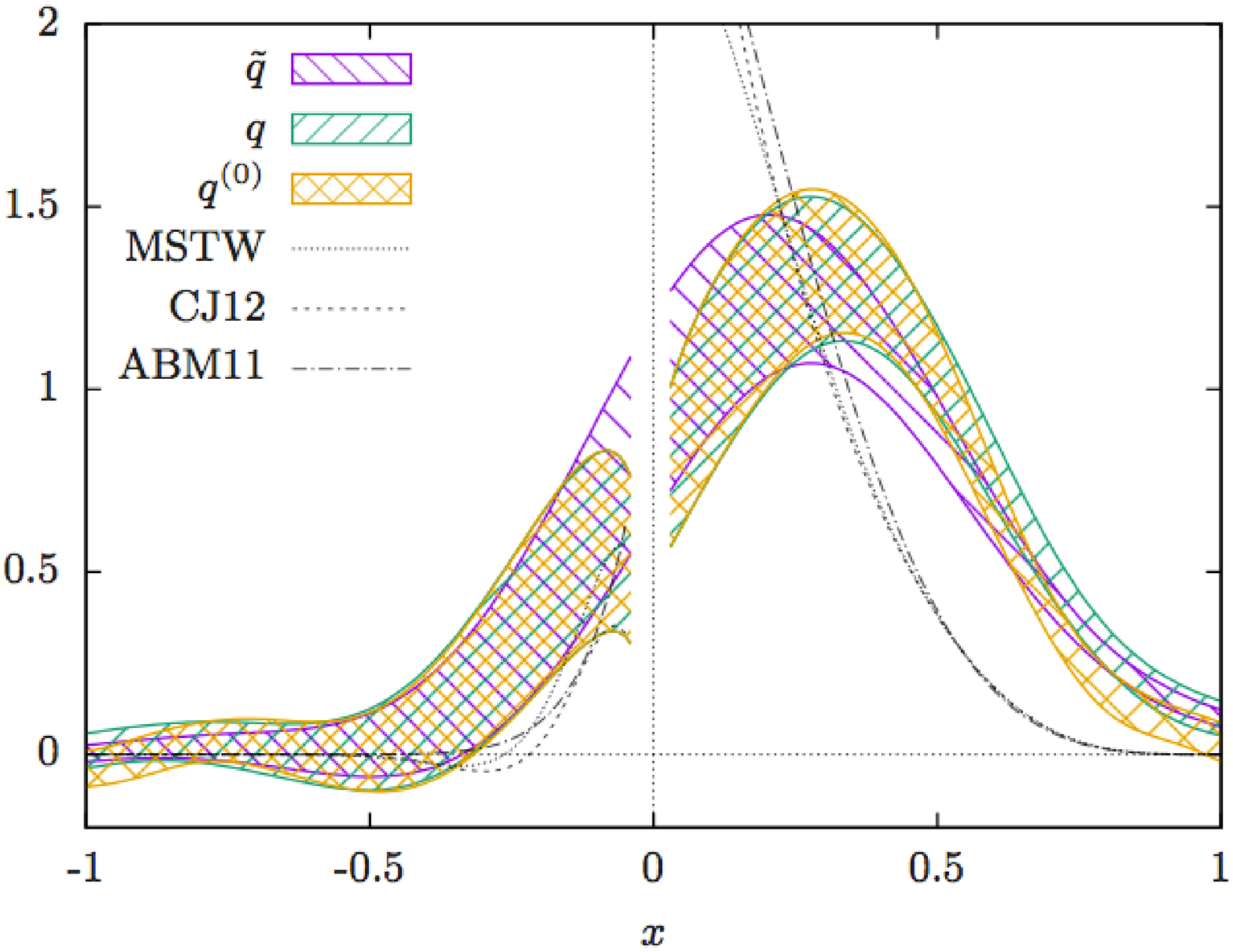}}
\vskip -0.4cm
\caption{Left: Results for the isovector $\tilde{q}$ upon one-loop and
  mass corrections for the momentum $P_z=4\pi/4$ and for 0, 2 and 5
  HYP smearing steps. Right: The quasi-distribution function
  $\tilde{q}$, the PDF without subtracting the mass correction $q$,
  and the final PDF, $q(0)$, shown for momentum
  $P_z=\frac{6\pi}{L}$. Various black lines show phenomenological
  results at 6.25 MeV$^2$ from MSTW~\cite{Martin:2009iq}
  (CJ12~\cite{Owens:2012bv}), ABM11~\cite{Alekhin:2012ig}.}
\label{figPDFsETMC}
\end{figure}

In the left panel of Fig.~\ref{figPDFsETMC} we show the isovector
$\tilde{q}$ for $P_z=\frac{4\pi}{L}$ and 0, 2, 5 HYP smearing
steps. As can be seen, the difference between the 0 and 2 steps is
more pronounced than the difference of 2 to 5 steps, indicating a
saturation of the smearing effect. These data correspond to the 1-loop
and nucleon mass corrected results, and both the real and imaginary
parts are taken into account. However, the difference between the
various HYP steps indicated that the proper renormalization will play
an important role, and the renormalized results are expected to agree
within statistical errors.

The physical quark distribution function $q$ can be extracted from
$\tilde{q}$ and then the mass corrections may be applied. This is
shown in the right panel of Fig.~\ref{figPDFsETMC} for momentum
$P_z=\frac{6\pi}{L}$. The authors find that while increasing the
momentum from $\frac{2\pi}{L}$ to $\frac{6\pi}{L}$, the peak of the
$u(x)-d(x)$ moves to smaller values of $x$, for the negative region,
the $\bar{d}(x)-\bar{u}(x)$ becomes very small for most of the $x<0$
region. The latter is in qualitative agreement with the behavior of
the antiquark distributions as extracted from phenomenological
analyses~\cite{Martin:2009iq,Owens:2012bv,Alekhin:2012ig}. The nucleon
mass corrections lead to a desirable decrease of the distributions in
the large $x$ region. Also, the mild oscillatory behavior in the large
$x$ region is due to the fact that the Fourier transformation was
performed over a finite interval, that is $−L/2 \le z \le +L/2$. 
\FloatBarrier

\section{Nucleon Spin}

In 1989 DIS experiments at CERN showed that only a small amount of the
proton spin was actually carried by the valence quarks. This was
called ``the proton spin crisis'', but since then our understanding
on the proton spin has evolved. We now know that both the gluons and
sea quarks are polarized and, thus, their contribution to the spin is
essential. It is also understood that a complete description of the
spin requires to take into account the non-perturbative structure of
the proton. Using the lattice QCD formalism one can provide
significant input towards understanding this open issue. The total
nucleon spin is generated by the sum of the quark orbital angular momentum
($L^q$), the quark spin ($\Sigma^q$) and the gluon angular momentum
($J^g$). The quark components are related to $g^q_A$ and the GFFs of
the one-derivative vector at $Q^2=0$
\vspace{-0.24cm}
\be
\frac{1}{2} = \sum_q \left(L^q + \frac{1}{2}\Delta\Sigma^q \right) + J^g\,,
\quad J^q = \frac{1}{2} \left( A_{20}^q + B_{20}^q  \right)\,,  \quad L^q=J^q-\Sigma^q \,,  \quad \Sigma^q = g_A^q\, ,
\label{J}
\ee 
\vskip -0.24cm
\noindent
where $L^q,\,\Delta\Sigma^q$ and $J^g$ are gauge invariant.
Since we are interested in the individual quark contributions to the
various components of the spin, one needs to consider the
disconnected contributions. 

The computation of disconnected diagrams using improved actions with
dynamical fermions became feasible over the last years and for the
proper renormalization of the individual quark and isoscalar
contributions one should take into account the singlet operator and
its proper renormalization (see Ref.~\cite{Constantinou:2014tga} for
further discussion).

\begin{figure}[!h]
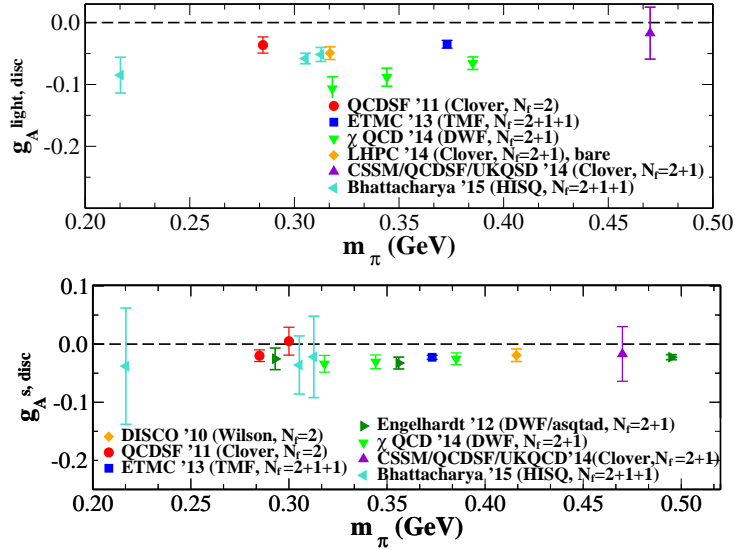

\cl{\includegraphics[scale=0.34]{./gA_disc_light_vs_mpi.eps}}
\vskip .1cm
\cl{\includegraphics[scale=0.34]{./gA_disc_strange_vs_mpi.eps}}
\vskip -0.4cm
\caption{Disconnected contributions for $g_A^q$ for the light (left) and
  strange (right) quark contributions.}
\label{fig15}
\end{figure}

A number of results have appeared recently where the disconnected loop
contributions to $g_A$ are evaluated as shown in Fig.~\ref{fig15}. We
observe a nice agreement among results using a number of methods both
for the light~\cite{QCDSF:2011aa,Abdel-Rehim:2013wlz,chiQCD14b,LHPC14,
QCDSF14,Bhattacharya:2015gma,ETMC14e}, as well as for the strange
quark contributions~\cite{Babich:2010at,QCDSF:2011aa,Abdel-Rehim:2013wlz,  
Engelhardt:2012gd, LHPC14,QCDSF14,Bhattacharya:2015gma,ETMC14e}. 
For $g_A^{light}$ we find $\sim10\%$ contributions compared to the
connected part that must be taken into account in the discussion of
the spin carried by quarks in the proton. These contributions are
negative and thus reduce the value of $g_A^q$.
\vskip -0.165cm
\FloatBarrier
\vskip -0.35cm
\begin{figure}[!h]
\cl{\includegraphics[scale=0.25,angle=90]{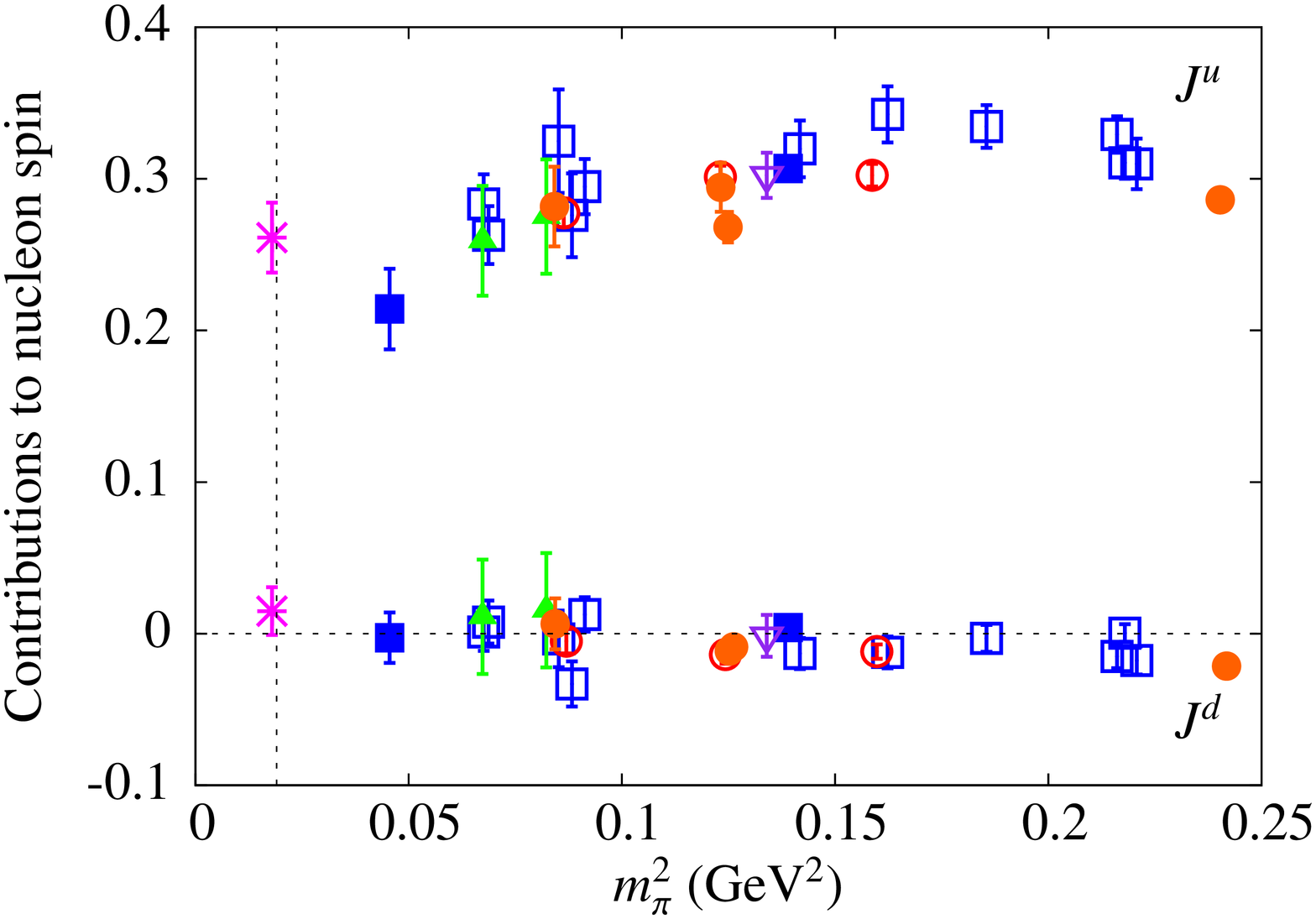}\,\,
    \includegraphics[scale=0.25,angle=90]{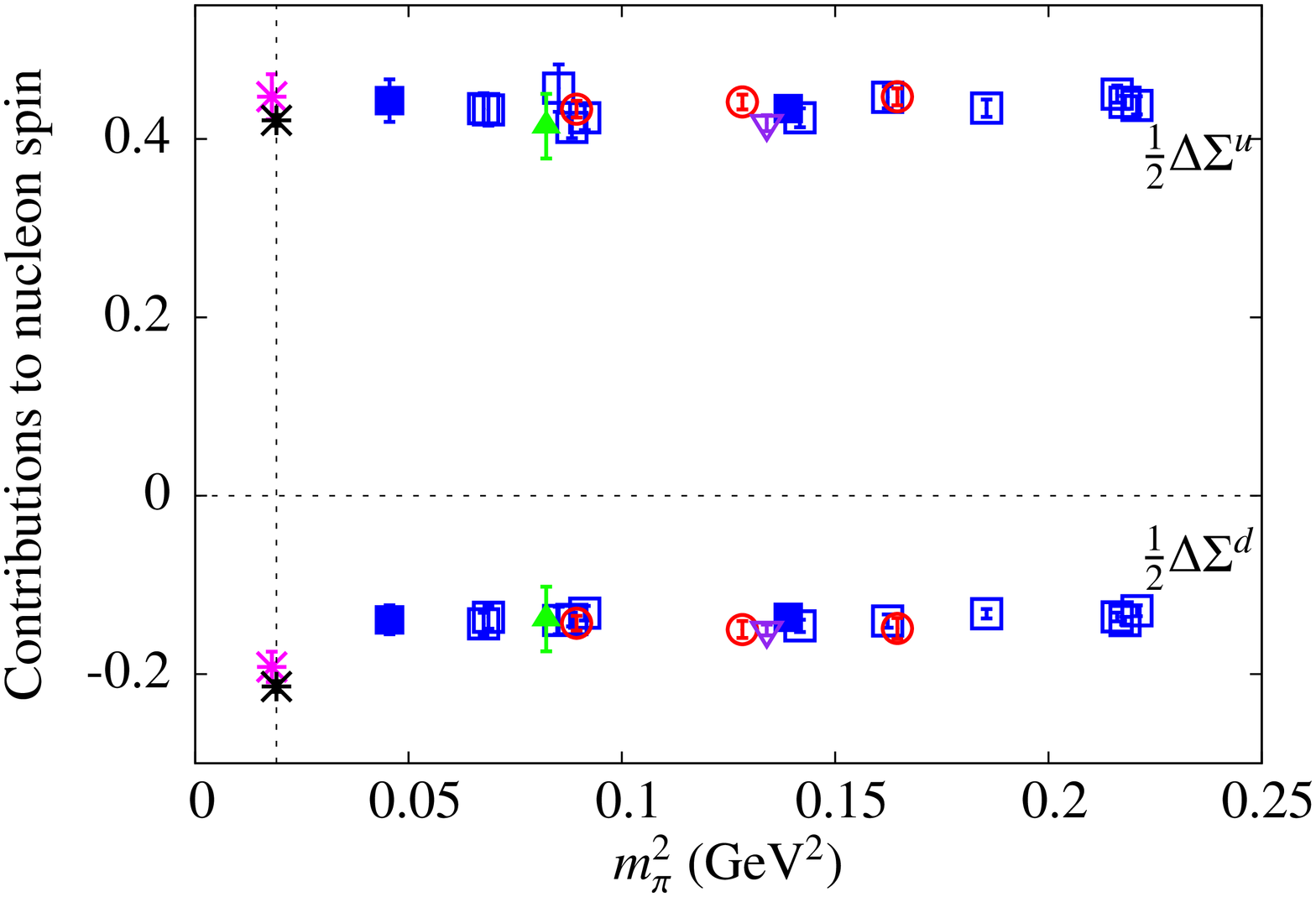}}
\vskip -0.45cm
\caption{The total spin, $J^q$, and the quark spin, $\Sigma^q$,
  carried by the up and down quarks. The lattice data correspond to:
$N_f{=}2{+}1$ DWF and DWF on asqtad (LHPC~\cite{Syritsyn:2011vk}),
$N_f{=}2$ Clover (QCDSF/UKQCD~\cite{Sternbeck:2012rw}),
$N_f{=}2$ TMF (ETMC~\cite{Alexandrou:2011nr})
$N_f{=}2{+}1{+}1$ TMF (ETMC~\cite{Alexandrou:2013joa})
$N_f{=}2$ TMF with Clover (ETMC~\cite{Abdel-Rehim:2015owa}). }
\label{fig16}
\end{figure}
\FloatBarrier
In Fig.~\ref{fig16} we show results for the u and d contributions to
the total spin, $J^q$. It is found that the u-quark exclusively
carries the spin attributed to the quarks in the nucleon since $J^d$
is consistent with zero for all pion masses and lattice discretization
schemes. The quark distribution to the intrinsic spin in also shown in
Fig.~\ref{fig16}. There is a nice agreement between the results at
the physical pion mass using TM fermions~\cite{Abdel-Rehim:2015owa} and the
experimental values for both the u- and the d-quarks. The disconnected
contributions have been neglected from most data except for one TMF
ensemble at $m_\pi=375$ MeV. The effect is shown by the shift of
the filled blue square -which ignores disconnected contributions- to
the violet triangle -which includes them. Although the effect is small,
it is larger than the statistical error and thus one needs to take
them into account. The lattice results thus corroborate the missing
spin contribution arising from the quarks. 

\section{Gluon moments of the nucleon}
\label{secGluonMoment}

To go further in our understanding for the structure of the nucleon
and the missing contributions to its spin we need to consider
contribution from the gluonic degrees of freedom. In this section, 
we will discuss lattice results that predict a sizeable
contribution of the gluons to the nucleon spin.
Investigation of the gluon distribution functions has also become of
high importance in major experimental facilities such as COMPASS
and STAR. Experimentally the gluon distributions can be determined from
the QCD evolution of the DIS and DIS measurements and a number of groups are
carrying out extended analyses.

While the quark moments have been studied
extensively~\cite{Abdel-Rehim:2015owa} there are only a few
computations for the gluon moments, mainly due to the bad 
signal-to-noise ratio, as well as the fact that there is a mixing with
the corresponding quark singlet operator. Here we discuss recent results for
the unpolarized gluon moment using different methods. This observable
can be evaluated by employing the following gluon operator
\be
{\cal O}_{\mu\nu}^g = -{\rm Tr}\left[ G_{\mu\rho} G_{\nu\rho}  \right],\,
\ee
from which one may extract the gluon moment by constructing
appropriate combinations of Dirac indices
\be
\langle N(p) |{\cal O}^L_b  | N(p) \rangle = 
\left( m_N + \frac{2}{3\,E_N}\vec{p}^2 \right)\,\langle x \rangle_g\,,\quad
{\cal O}^L_b \equiv {\cal O}^L_{44} - \frac{1}{3} \sum_{j=1}^3 {\cal O}^L_{jj}.\,
\label{RGluon}
\ee
 The lattice discretization of
the gluon operator is denoted by ${\cal O}^L_b $ and can be expressed
in terms of plaquettes
\be
{\cal O}^L_b = \frac{4}{9}\frac{\beta}{a}\sum_x\left(\sum_i\mbox{tr}_c[U_{i4}(x,t)]-\sum_{i<j}\mbox{tr}_c[U_{ij}(x,t)]\right)\, .
\ee
The advantage of this operator is that the gluon moment,  $\langle x
\rangle_g$, can be extracted directly from lattice data at zero
momentum transfer, as can be seen from the rhs of
Eq.~(\ref{RGluon}). However, the fact that terms of similar magnitude
are subtracted leads potentially to a noisy signal.

A direct computation of the gluon moment is related to the following
ratio at zero momentum transfer
\be
  \frac{\left\langle [N(t)N(0)]_{p=0} \mathcal
    B(\tau)\right\rangle}{\langle N(t)N(0)_{p=0} \rangle}
  \stackrel{0\ll \tau \ll t}=m_N \langle x \rangle_g\,,
\ee
which comprises of  disconnected diagrams only.
The three-point function can, thus, be written as a
product of nucleon two-point functions and the gluon operator. 
Although disconnected contributions are notoriously difficult and
noisy,  applying smearing to the gauge links
 in the gluon operator improves the quality of the 
signal. This was demonstrated in Ref.~\cite{Alexandrou:2013tfa} using
$N_f{=}2+1+1$ twisted mass fermions at $m_\pi\sim 373$ MeV. The
authors test both HYP and stout smearings and find a significant
reduction of the noise-to-signal ratios after five steps for the HYP
smearing, and ten steps for the stout smearing. 
\vspace*{-0.25cm}
\begin{figure}[!h]
\cl{\includegraphics[scale=0.4]{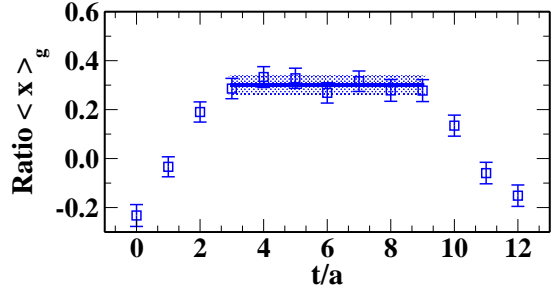}}
\vskip -0.1cm
\caption{The ratio for $\langle x \rangle_g$ for $N_f{=}2+1+1$ twisted
  mass fermions at $m_\pi\sim 375$ MeV after 10 iterations of stout
  smearing steps.}
\label{figRGluon}
\end{figure}
\FloatBarrier
\vspace*{-0.15cm}
An example is shown in
Fig.~\ref{figRGluon} for the ratio leading to $\langle x\rangle_g$ after
10 iterations of stout smearing steps.
A challenge with such a computation is that to obtain physical results
for $\langle x \rangle_g$, the lattice matrix element needs to be
renormalized. Since the gluon operator is singlet it mixes
with the quark momentum fraction $\langle x \rangle_q$, as well as
with other operators that are: (a) gauge invariant, or (b)
BRS-variations, or (c) vanish by the equations of motion. However, in
physical matrix elements the mixing with the operators (a)-(c)
vanishes and the mixing reduces to a $2\times2$ matrix, that is
\bea
   \left( \begin{array}{c}
             \langle x \rangle_g^{\msbar}(\mu) \\
             \sum_q \langle x \rangle_q^{\msbar}(\mu) \\
          \end{array}
   \right)
   = \left( \begin{array}{cc}
               Z_{gg}^{\msbar}(\mu)&
               Z_{gq}^{\msbar}(\mu)\\
               Z_{qg}^{\msbar}(\mu)&
               Z_{qq}^{\msbar}(\mu)\\
           \end{array}
     \right) \, \left( \begin{array}{c}
                          \langle x \rangle_g \\
                          \sum_q \langle x \rangle_q \\
                       \end{array}
                \right) \,,
\eea
where $\mu$ is the renormalization scale, usually set to 2~GeV.
Note that, in the quenched approximation the mixing matrix simplifies
considerably since both $Z_{gq}$ and $Z_{qg}$ become $1-Z_{qq}$
and $1-Z_{gg}$, respectively. For the renormalization of 
$ \langle x \rangle_g$ the relevant matrix elements are  
$Z_{gg}$ and $Z_{qq}$ and the relation to the bare quark and gluon
moments is
\be
  \langle x \rangle_g^{\msbar} =
  Z_{gg}^{\msbar}\langle x \rangle_g +
  Z_{gq}^{\msbar}\sum_q \langle x \rangle_q\, .
\ee
Due to the mixing and the involvement of disconnected contributions,
 an appropriate renormalization scheme
to extract the multiplicative renormalization functions and the mixing
coefficients non-perturbatively is a difficult task. As a first step we thus use  perturbation theory to compute the
elements of the mixing matrix. One of the advantages of the perturbative
calculation~\cite{GluonPert} is that the results can be computed
directly in the $\msbar$ scheme without an intermediate RI-type step. 
Since the gauge links of the operator are smeared for signal
improvement, an equivalent procedure is also followed in the
perturbative calculation in order to match the non-perturbative
calculation of $ \langle x \rangle_g$. In the calculation of
Ref.~\cite{GluonETMC} the preferred smearing is stout since it is
analytically defined in both the perturbative and non-perturbative
evaluations. Note, however, that in the perturbative computation of
the renormalization functions introduction of smearing  increases
the number of algebraic expression, which explodes as the stout
smearing steps increase. Currently, the computation is performed to 2
smearing steps, which already involved millions of terms. The smearing
parameter is chosen to be small, and thus, more levels of smearing is
expected to bring in a very small effect, due to the polynomial
dependence on the smearing parameter. For the work of Ref.~\cite{GluonETMC} 
the renormalized matrix element at $m_\pi=373$ MeV in the $\msbar$ at 2
GeV is found to be: $ \langle x \rangle_g=0.309(25)$.

\begin{figure}[!h]
\cl{\includegraphics[scale=0.4]{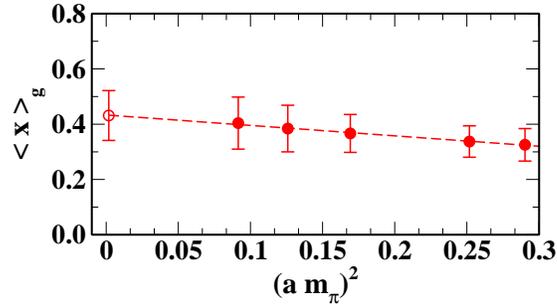}}
\vskip -0.1cm
\caption{$\langle x \rangle_g$ for $N_f{=}0$ clover
fermions~\cite{Horsley:2012pz} as a functions of the pion mass squared,
$(a\,m_\pi)^2$. The open circle corresponds to the chirally
extrapolated result.}
\label{gluonQCDSF}
\end{figure}

An alternative approach to extract the matrix elements of the gluon
operator utilizes the Euclidean form of the Feynman-Hellman
theorem. In this methodology an
operator $\lambda {\cal O}$ is introduced into the total QCD action,
and the matrix element of the operator can be extracted from the 
derivative of the energy of the state with respect to $\lambda$
\be
\frac{\partial E_N(\lambda)}{\partial\lambda} = ( :\frac{\partial
  \hat{S}(\lambda)}{\partial \lambda}:)_{N(p)N(p),\lambda}\,,
\ee
where $:...:$ denotes the subtraction of the vacuum expectation value
of the operator. By combining the above equations with the
continuum decomposition expression, one can extract the gluon moment
at zero momentum transfer
\be
\langle x \rangle_g = \frac{2}{3m_N}\frac{\partial m_N}{\partial
\lambda}{\big |_{\lambda=0}}\,.
\ee
Note that the Feynman-Hellman methodology requires production of new
gauge ensembles for each value of the $\lambda$ parameter (and each
operator), which is computationally costly, especially for $m_\pi$
close to the physical value. This methodology was applied in
Ref.~\cite{Horsley:2012pz} in the quenched approximation for 
clover fermions at ensembles corresponding to several values of $m_\pi$ so
that the extrapolation to the chiral limit can be taken, as shown in
Fig.~\ref{gluonQCDSF}. The extrapolated value is $\langle x \rangle_g
= 0.43(7)(5)$, which, despite being quenched, it is close to the value
of ETMC using dynamical twisted mass fermions~\cite{GluonETMC}. 
\vspace*{-0.25cm}
\FloatBarrier

A different direction for the computation of not only the gluon, but
also the quark moments relies on a complete gauge-invariant
decomposition of the nucleon spin (see Eq.~(\ref{J})) in terms of the
quark spin, the quark orbital angular momentum, and the glue angular
momentum operators, as defined from the symmetric energy-momentum
tensor. Thus, instead of computing directly $J_q$ and $J_g$ from the
explicit definitions
\be
\vec{J}_q = \frac{1}{2} \vec\Sigma_q + \vec{L}_q  =  \int d^3x \,  \bigg{[} \frac{1}{2}\, \overline\psi\,\vec{\gamma}\,\gamma^5 \,\psi
 + \psi^\dag \,\{ \vec{x} \times (i \vec{D}) \} \,\psi \bigg{]} \,,\quad
\vec{J}_g = \int d^3x \,\bigg{[} \vec{x} \times ( \vec{E} \times  \vec{B} )\bigg{]}\,,
\ee
one can calculate them from the energy-momentum tensor
\be
J_{q,g}^i = \frac{1}{2}\,\epsilon^{ijk}\,\int \, d^3x\, \left(\mathcal{T}_{q,g}^{0k}\, x^j  - \mathcal{T}_{q,g}^{0j}\, x^k\right)\, ,
\ee
which in Euclidean space the quark and gluon operators are
\bea
  {\mathcal T}_{\{4i\}q}^{(E)}
  &=&  (-1)\, \frac{i}{4}\displaystyle\sum_f \overline {\psi}_f
  \left[
    \gamma_4 \stackrel{\rightarrow} D_i
    + \gamma_i \stackrel{\rightarrow} D_4
    -  \gamma_4 \stackrel{\leftarrow} D_i
    - \gamma_i \stackrel{\leftarrow} D_4
    \right] \psi_f , \\
  {\mathcal T}_{\{4i\}g}^{(E)}
  &=&  (+i)\, \bigg[-\frac{1}{2} \displaystyle\sum_{k=1}^3 2\,
    \mbox{Tr}^{\rmsmall{color}} \left[G_{4k}\, G_{ki} + G_{ik}\,
      G_{k4} \right]\bigg] .
\eea
The complete calculation of the quark and glue momenta and angular
momenta on a quenched lattice for Wilson fermions has been presented
in Ref.~\cite{Deka:2013zha}, including both connected and disconnected
insertions for the quark contributions. Three ensembles have been
employed with $m_\pi=$478, 538, 650 MeV. The overlap operator is used
for the gauge field tensor, which leads to less noisy results
than that from usual gauge links. Details on the computation can be
found in Ref.~\cite{Deka:2013zha}. Regarding the renormalization of
the operators, the authors use  sum rules to define renormalization
conditions on the lattice. This results from the fact that although
the momentum and angular momentum fractions of the quark and glue are
renormalization scale and scheme dependent individually, their sums are
not because the nucleon total momentum and angular momentum are
conserved. One thus obtains
\be
J_{q,g} + \frac{1}{2} Z_{q,g}^L \left[T_1(0) + T_2(0)\right]_{q,g} \,,\quad
\langle x\rangle_{q,g} = Z_{q,g}^L T_1(0)_{q,g}\,,
\ee
and
\bea
\langle x\rangle_{q} +   \langle x\rangle_{g}\, =\, Z_q^L T_1 (0)_q
+ Z_g^L T_1 (0)_g  &=& 1 \,, \\
J_q+ J_g\, =\, \frac{1}{2}\,  \bigg\{
Z_q^L \left[ T_1 (0) + T_2 (0) \right]_q + Z_g^L \left[ T_1 (0)  +
  T_2 (0) \right]_g
\bigg\} &=& \frac{1}{2} \,,
\eea
which also leads to
\be
Z_q^L T_2 (0)_q + Z_g ^L T_2 (0)_g =0\,,
\ee
providing sufficient conditions and cross-checks to obtain the
renormalization functions.
\begin{figure}[!h]
\cl{\includegraphics[scale=0.35,angle=-90]{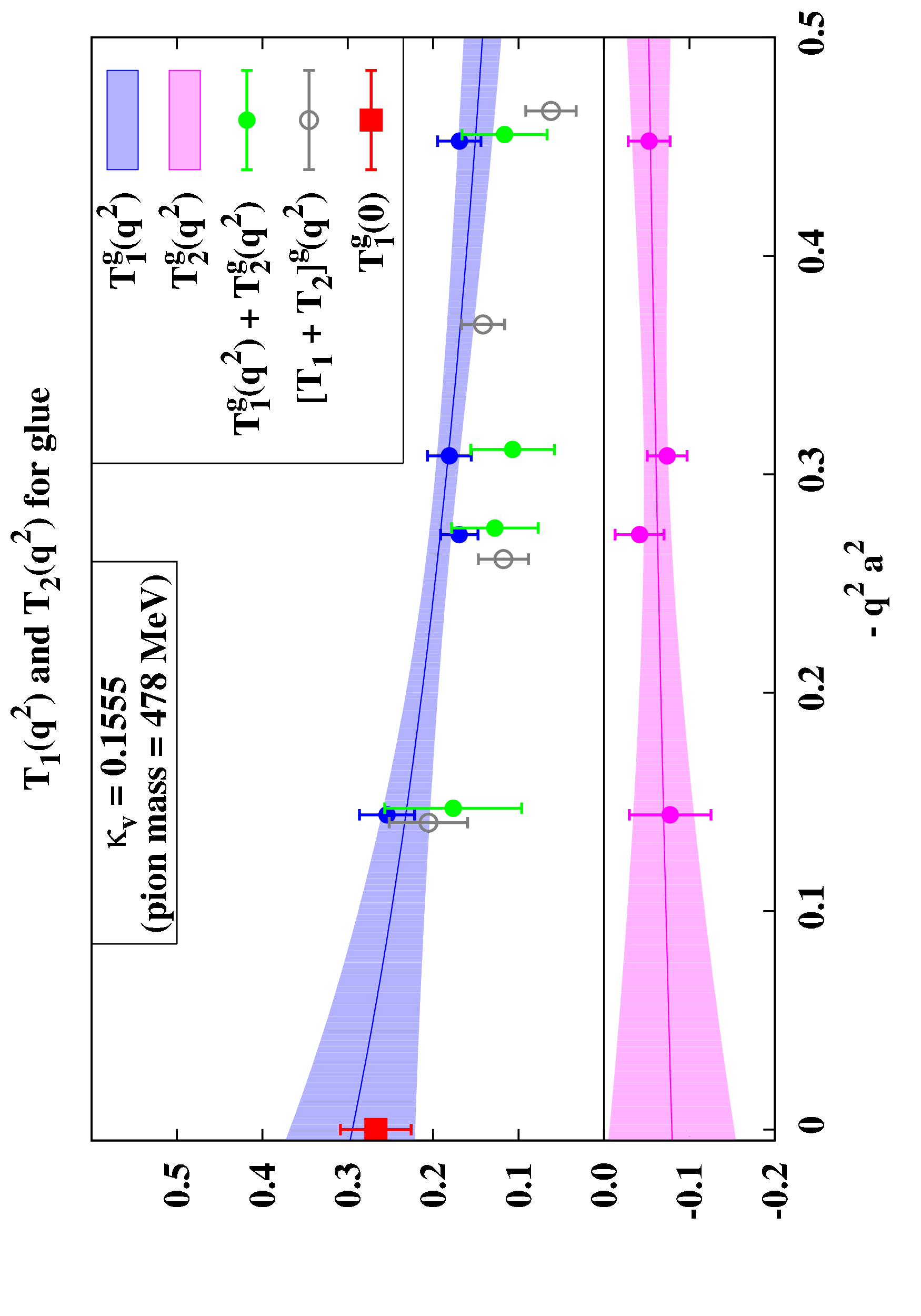}}
\vskip -0.1cm
\caption{$T_1$, $T_2$ for gluons and angular momentum $J_g$ using
quenched Wilson fermions~\cite{Deka:2013zha}.} 
\label{gluonxQCD}
\end{figure}
\FloatBarrier
In Fig.~\ref{gluonxQCD} we show results for the gluon contributions
corresponding to 478 MeV. The extrapolated value for the gluon
unpolarized moment is $\langle x \rangle_g=0.313(56)$, which is
compatible with the results presented above for both quenched
\cite{Horsley:2012pz} and dynamical fermions~\cite{GluonETMC}.

\section{Neutron Electric Dipole Moment}

 There is recently a major activity  in LQCD
computation of the neutron electric dipole moment (EDM),
$\vec{d}_N$, which we review in this section.
 A non-zero EDM indicates the violation of parity
($P$) and time ($T$) symmetries, and consequently of $CP$, probing
physics BSM~\cite{Pospelov:2005pr}. So far, no finite neutron EDM (nEDM) has been
reported and current bounds are still several orders of magnitude
below what one expects from $CP$-violation induced by weak
interactions. Several experiments are under way to improve the upper
bound on the nEDM value, with the best experimental upper limit 
being~\cite{Helaine:2014ona,Baker:2006ts,Baker:2007df}
\be
\vert \vec{d}_N \vert  < 2.9 \times 10^{-13} e \cdot {\rm fm} \ (90\% \ {\rm CL})\,.
\label{eq:nEDM_smaller_experimental_upperbound}
\ee 

To investigate theoretically a finite nEDM, we add to the
$CP$-conserving QCD Lagrangian density a $CP$-violating
interaction term, proportional to the topological charge, $q$
\bea
  {\cal L}_{\rm QCD} \left( x \right) = \frac{1}{2 g^2} 
{\rm Tr} \left[ G_{\mu \nu} \left( x \right) G_{\mu \nu} \left( x \right) \right] + 
\sum_{f} {\overline \psi}_{f} \left( x \right) (\gamma_{\mu} D_{\mu} +m_f) \psi_{f}\left( x \right)\,
- i \theta q \left( x \right)\,,\\
  q \left( x \right) = \frac{1}{ 32 \pi^2} \epsilon_{\mu \nu \rho \sigma} 
{\rm Tr} \left[ G_{\mu \nu} \left( x \right) G_{\rho \sigma} \left( x \right) \right]\,,\hspace*{3cm}
\eea
where $\psi_f$ denotes a fermion field of flavor $f$ with bare mass
$m_f$ and $G_{\mu \nu}$ is the gluon field tensor. The so called
$\theta$-parameter controls the strength of the $CP$-breaking, and the
addition of the $CP$-violating term leads to a non-zero value for
nEDM. The $\theta$-parameter can be taken as a small continuous
parameter allowing a perturbative expansion and only keep first
order contributions in $\theta$. This is in accordance to effective
field theory calculations (see references within~\cite{Alexandrou:2015spa})
%~\cite{Pich:1991fq,Borasoy:2000pq,Hockings:2005cn,
%Narison:2008jp,Ottnad:2009jw,deVries:2010ah,Mereghetti:2010kp,deVries:2012ab,Guo:2012vf}
that give a bound of the order 
$\theta \lesssim {\cal O} \left(10^{-10} - 10^{-11} \right)$.
\begin{figure}[!h]
\cl{\includegraphics[scale=0.36]{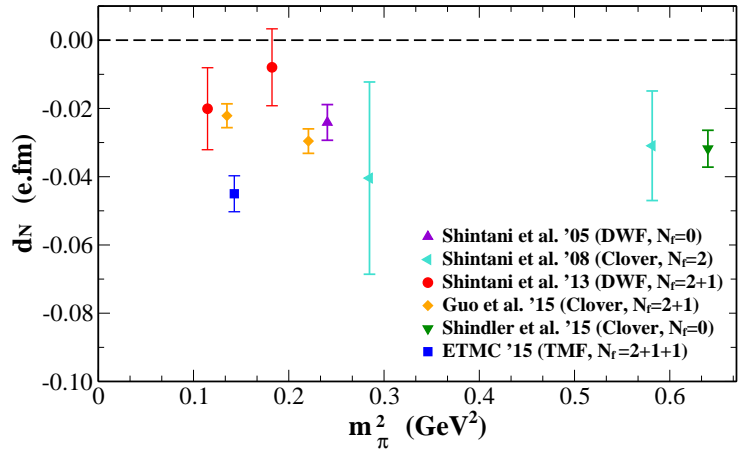}}
\caption{The nEDM versus $m_\pi^2$ for: a) $N_f{=}2{+}1{+}1$ twisted
mass fermions~\cite{Alexandrou:2015spa} (blue square) corresponding to a
weighted average using different methods for extracting $F_3(0)$), 
b) $N_f{=}0$ DWF~\cite{Shintani:2005xg} (magenta upward triangles),
$N_f{=}2{+}1$ DWF~\cite{Shintani:2014zra} (red circles) and 
$N_f{=}0$ clover fermions~\cite{Shindler:2015aqa} (green downward
triangles) by extracting the $CP$-odd $F_3(Q^2)$ and fitting its
$Q^2$-dependence, c) $N_f{=}2$ Clover fermions~\cite{Shintani:2008nt}
(turquoise left triangles) obtained using a background electric field,
d) $N_f{=}2{+}1$ clover fermions~\cite{Guo:2015tla} (orange diamonds)
by implementing an imaginary $\theta$.}
\label{nEDM}
\end{figure}
\FloatBarrier

The quantity, which is of  interest is  the nEDM,
$\vec{d}_N$, which at leading order of $\theta$ and in momentum space
is given by~\cite{Jarlskog:1988}
\bea
\vert \vec{d}_N \vert =  \theta \lim_{Q^2 \to 0} \frac{\vert F_3(Q^2) \vert}{2 m_N}\,,
\label{eq:dN}
\eea
where $m_N$ denotes the mass of the neutron, $Q^2{=}-q^2$ the
four-momentum transfer in Euclidean space ($q{=}p_f-p_i$) and
$F_3(Q^2)$ is the $CP$-odd form factor. In a theory
with $CP$ violation we can, therefore, calculate the electric dipole
moment by evaluating the zero momentum transfer limit of the
$CP$-odd form factor. However, the $CP$-violating matrix element gives
access to $Q_k F_3(Q^2)$ and not to $F_3(Q^2)$ alone, hindering a
direct evaluation of $F_3(0)$. Details on different methods for the
extraction of $F_3(Q^2)$ can be found in Ref.~\cite{Alexandrou:2015spa}.

Besides extracting the nEDM from the $CP$-odd form factor
\cite{Shintani:2005xg,Shintani:2014zra,Shindler:2015aqa,Alexandrou:2015spa}, 
there are alternative methods to compute the nEDM, such are the
implementation of an imaginary $\theta$~\cite{Guo:2015tla}, or with an
application of an external electric field and measuring the associated
energy shifts \cite{Shintani:2008nt}. A collection of lattice results
are displayed in Fig.~\ref{nEDM}, using different methods for
obtaining $d_N$, as well as, different definition of the topological
charge. We note that the results of Ref.~\cite{Shindler:2015aqa} have
been extrapolated to the continuum limit.

\section{Summary and Challenges}

Lattice QCD has entered a new era in terms of simulations with the
light quark masses fixed to their physical value. This is due to major
improvements in algorithm and techniques coupled with increase in the
computational power. However, many challenges lie ahead: 
development of appropriate algorithms to reduce the statistical errors
at reduced cost and addressing systematic uncertainties in order to
compute accurately observables that reproduce experimental data or can
probe beyond the standard model physics.

For hadron structure, simulations at different lattice spacings and
larger volumes are crucial for a proper study of lattice artifacts in
order to provide reliable results at the continuum limit. Such studies 
require an accuracy, which is difficult to achieve with standard methods.
Noise reduction techniques are, thus, essential in order to settle
some of the long-standing discrepancies reviewed in this talk.
Similarly techniques developed for the computation of disconnected quark
loop diagrams, such as the truncated solver method~\cite{Bali:2009hu}
need to be improved since they become inefficient at the physical
point~\cite{ETMC14e}. Thus, new ideas will be needed to compute
disconnected contributions to  hadron structure to an accuracy of a few
percent. Utilization of new computer architectures such as GPUs has
proved essential for the evaluation of disconnected diagrams and this
is a  direction that we will further   pursue in the future.  

Other open issues such as the nucleon spin need the evaluation
of  quantities that are challenging to compute,  such as  gluonic contributions. As
discussed in this review, there are several challenges in the
computation of the gluon moments, including increased gauge noise and
mixing with other operators. Perturbation theory has been utilized in
order to successfully compute the multiplicative renormalization and
disentangle the operator mixing. 

Evaluation of the nucleon matrix elements for the electromagnetic
current in a theory with a $CP$-violating term in the Lagrangian,
yields the value of the neutron electric dipole moment from
first principles. Despite its difficulties, such a calculation can
guide planned experiments.

Despite the challenges of LQCD calculations, simulations at the 
physical point have eliminated one of the systematic uncertainties
that was inherent in all lattice calculations in the past, that is the
difficulty to quantify systematic error due to the chiral extrapolation.
Calculating observables directly at the physical point   holds the promise of resolving
discrepancies on benchmark quantities like $g_A$ and reliably compute
quantities relevant for revealing possible new physics BSM. 

\bigskip
{\bf {Acknowledgments:}} I would like to thank the organizers for
their invitation to participate and for providing partial financial
support. I also want to thank my collaborators for fruitful
discussions.

\end{document}